\documentclass[lettersize,journal]{IEEEtran}
\usepackage{amsmath,amsfonts}
\usepackage{algorithmic}
\usepackage{algorithm}
\usepackage{array}
\usepackage[caption=false,font=normalsize,labelfont=sf,textfont=sf]{subfig}
\usepackage{textcomp}
\usepackage{stfloats}
\usepackage{url}
\usepackage{verbatim}
\usepackage{graphicx}
\usepackage{cite}
\begin{document}

\title{Problem-Structure-Informed Quantum Approximate Optimization Algorithm for Large-Scale Unit Commitment with Limited Qubits}

\author{Jingxian Zhou, Ziqing Zhu,~\IEEEmembership{Member,~IEEE,} Linghua Zhu, Siqi Bu,~\IEEEmembership{Senior Member,~IEEE}
}

\maketitle

\begin{abstract}
As power systems expand, solving the Unit Commitment Problem (UCP) becomes increasingly challenging due to the dimensional catastrophe, and traditional methods often struggle to balance computational efficiency and solution quality. To tackle this issue, we propose a problem-structure-informed Quantum Approximate Optimization Algorithm (QAOA) framework that fully exploits the quantum advantage under extremely limited quantum resources. Specifically, we leverage the inherent topological structure of power systems to decompose large-scale UCP instances into smaller subproblems, each solvable in parallel by limited number of qubits. This decomposition not only circumvents the current hardware limitations of quantum computing but also achieves higher performance as the graph structure of the power system becomes more sparse. Consequently, our approach can be readily extended to future power systems that are larger and more complex.
\end{abstract}

\begin{IEEEkeywords}
Unit Commitment Problem, Quadratic Unconstrained Binary Optimization, Quantum Approximate Optimization Algorithm.
\end{IEEEkeywords}

\section{Introduction}
\IEEEPARstart{T}{he} unit commitment problem (UCP) aims to achieve cost minimization while maintaining a reliable power supply through optimal scheduling (on/off) decisions of generators. Existing solution approaches primarily fall into three categories: exact algorithms, heuristic methods, and reinforcement learning techniques. Exact algorithms \cite{ref10} (e.g., Mixed-Integer Linear Programming, branch-and-bound) can guarantee an optimal solution or optimality bounds, but often require significant computational resources and may become impractical for large-scale problems. Heuristic methods \cite{ref11} (e.g., Genetic Algorithms, Particle Swarm Optimization, Tabu Search) are more scalable and flexible but do not guarantee global optimality, as they trade solution quality for reduced computation time. Reinforcement learning \cite{ref12} techniques model scheduling decisions as a Markov Decision Process, enabling adaptive policy learning over time; however, they typically necessitate substantial training efforts, hyper-parameter tuning, and computational overhead to converge to effective solutions.

In contrast to conventional methods, quantum computing demonstrates unique advantages in terms of computational speed, scalability, and guaranteed optimality, through the superposition principle that enables qubits to represent multiple states concurrently. By exploiting quantum entanglement, such systems achieve instantaneous correlation between computational trajectories, allowing parallel processing paths to collectively contribute to solution refinement. This capability translates into exponential acceleration potential for specific computational tasks with inherent parallel structures. Notably, the Quantum Approximate Optimization Algorithm (QAOA)~\cite{ref9} emerges as a promising candidate for combinatorial optimization problems like UCP. This hybrid algorithm applies alternating operators between problem-specific and mixing Hamiltonians, with parameter optimization conducted through classical computation loops to search optimal configurations.

However, while UCP typically involves hundreds to thousands of nodes, current quantum hardware faces scalability constraints due to limited qubit availability. This mismatch renders direct QAOA implementation theoretically inadequate for practical UCP requirements. To bridge this gap, we develop an problem-structure-informed QAOA (PSI-QAOA) framework tailored for UCP's structural demands. The proposed methodology achieves large-scale UCP solutions through a novel problem decomposition method informed by the structure of power system topology, thereby solving each sub-problem in parallel by limited number of qubits.

The remainer of this paper is organized as follows. Section II formulates the UCP. Section III introduces the proposed PSI-QAOA. Section IV demonstrates the experimental results, and Section V concludes this paper.

\begin{figure*}[!t]
\centering
\includegraphics[width=1\linewidth]{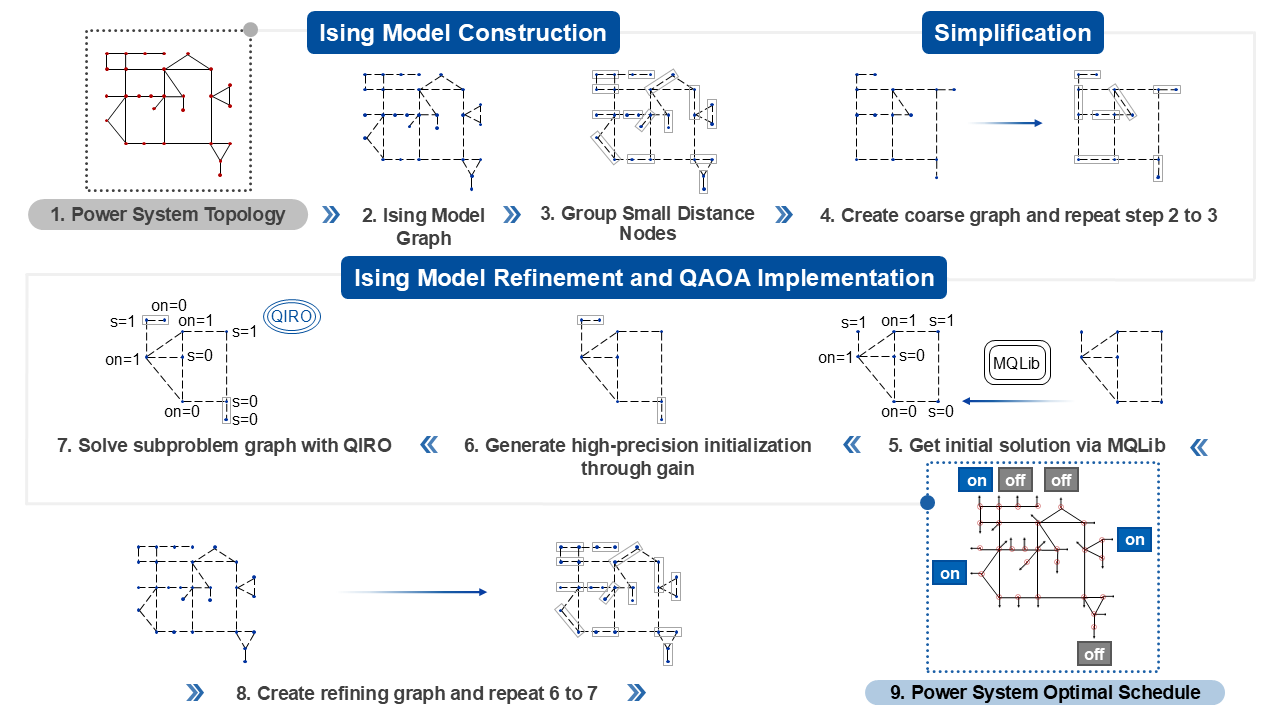}
\caption{Detailed description of PSI-QAOA. The power system topology (Step 1) with red dots as units and solid lines as circuits is mapped to a QUBO based Ising model (Step 2), where blue dots represent decision variables and dashed lines indicate node relationships based on the objective function and constraints. The original graph (Step 3) is simplified to a coarsest graph (Step 4), which is solved by MQLib for an initial solution (Step 5). This is followed by refinement of the Ising model with gain via QIRO (Step 6 to Step 8), and finally mapping back to the original power system optimal schedule (Step 9).}
\label{fig1}
\end{figure*}

\section{Problem Formulation}
The first step of solving UCP using QAOA is to transform the UCP into the Quadratic Unconstrained Binary Optimization (QUBO), as represented by (\ref{eq1}) \cite{ref1}:  
\begin{equation}
\label{eq1}
\begin{aligned}
  & \min \sum_{t\in T}{\sum_{i\in I}{\left( C_{i}  \max p_{i} on_{t,i} + H_{i}  s_{t,i} \right)}} \\ 
  & + A  \left( \sum_{t\in T} \left( \sum_{i\in I} \max p_{i}  on_{t,i} - D_{t} \right) \right)^{2} \\ 
  & + B  \sum_{t\in T} \sum_{i\in I} \left( on_{t,i} (1 - on_{t-1,i}) + 2 s_{t,i}  (on_{t-1,i} - on_{t,i}) + s_{t,i} \right) \\ 
  & + C  \sum_{t\in T} \sum_{i\in I} \left( s_{t,i}  \min up_{i} - \sum_{\tau =t}^{t-1+\min up_{i}} s_{t,i}  on_{\tau,i} \right) \\ 
  & + D \sum_{t\in T} \sum_{i\in I} \left( \sum_{\tau =t+1-\min down_{i}}^{t} \left( s_{t,i} + on_{t-1,i} - on_{t,i} \right) on_{\tau,i} \right)
\end{aligned}
\end{equation}
where the binary variables ${{s}_{t,i}}$ and ${{on}_{t,i}}$ denote the start and operating states of unit $i$ at time $t$. ${{C}_{i}}$ and  ${{H}_{i}}$ represent the linear production-dependent cost and fixed startup cost of unit $i$, while $\max {{p}_{i}}$ is its maximum power output. $T$ and $I$ denote the time period and unit set, respectively, with $A$, $B$, $C$, $D$ as penalty factors for constraint violations.

\section{Proposed Methodology}
In this section, we propose the PSI-QAOA method for solving the above QUBO problem. The process is comprised of three stages: Ising model construction, Ising model simplification, Ising model refinement and QAOA implementation. The framework is illustrated in Fig. \ref{fig1} with pseudo code provided in Algorithm \ref{alg1}.

\subsection{Ising Model Construction and Simplification}
After the UCP problem being re-formulated as the QUBO form (1), it is further represented as the Ising model graph to be processed by QAOA. Specifically, nodes in the graph correspond to decision variables (the start state and the operating state of the unit), while edges represent the interactions between these variables (network constraints). Then, the nodes are initially positioned on the unit sphere or hypercube at random, and multiple iterations are executed to maximize the weighted distance between adjacent nodes. This process can be expressed as (\ref{eq2}) \cite{ref2}.
\begin{equation}
\label{eq2}
\forall i\in {{N}_{l}},p_{i}^{t+1}\leftarrow \max \sum\limits_{j\in n(i)}{{{w}_{ij}}{{\left\| p_{i}^{t}-p_{j}^{t} \right\|}_{2}}}
\end{equation}
where ${{N}_{l}}$ denotes the node set of the $l$th level, and $p_{i}^{t}$ represents the position of node $i$ at iteration $t$. $n(i)$ represents the set of neighbors of node $i$, while ${{w}_{ij}}$ represents the weight of the edge $ij$. Subsequently, this graph is simplified by utilizing embedding, pairing nodes with their nearest unpaired neighbors using a K-D tree. This process is repeated until the coarsest graph that is smaller than the set minimum size $m$ is achieved. Intuitively, the purpose of this step is to effectively reduces the complexity of the system by simplifying the network, similar to the process of aggregating the nodes of a large power grid into a smaller, more manageable subsystem. 

\subsection{Ising Model Refinement and QAOA Implementation}
In this stage, we firstly solve the simplified Ising graph model with MQLib's BURER2022 heuristic \cite{ref3}. The required qubits is decremented by half each time according to the number of nodes until it falls below the set minimum size $m$. This initial solution further undergoes fidelity-preserving interpolation via operator $R: N_{l+1} \to N_l$ where $x_i = x_{R(i)},\ \forall i\in N_{l+1}$, generating high-precision initialization for subsequent refinement. 

The refinement process prioritizes the restoration of the most critical nodes, where the criticality of the nodes is explicitly defined by the gain metric $g(i)$ in (\ref{eq3}) \cite{ref2}. This gain quantifies each node's impact on the system energy through:
\begin{equation}
\label{eq3}
\forall i\in {{N}_{l}},g(i)\leftarrow \sum\limits_{j\in n(i)}{{{w}_{ij}}{{(-1)}^{2{{x}_{i}}{{x}_{j}}-{{x}_{i}}-{{x}_{j}}}}}
\end{equation}
During iterative processing, the algorithm first ranks all nodes by their absolute gain values $|g(i)|$, then selects $H$ nodes with the highest gains to construct the subproblem graph. These subproblems are subsequently solved through the QAOA Implementation stage, in which the Quantum Informed Recursive Optimization (QIRO) method is employed to enhance the performance of QAOA by utilizing a recursive approach to optimize the parameters\cite{ref5}.

\begin{algorithm}[H]
\caption{PSI-QAOA}
\label{alg1}
\textbf{Input:} Graph $G_0$, min size $m$, subproblem size $H$ \\
\textbf{Output:} Solution $S$
\begin{algorithmic}[1]

\STATE \textbf{Coarsening Stage}
\WHILE{$|N_l| > m$} 
    \STATE $M_l \gets \text{matrix}(G_l)$, embed $G_l$ via (\ref{eq2})
    \STATE Pair node with nearest neighbor, build $P$ matrix
    \STATE $G_{l+1} \gets P^\top M_l P$, $l \gets l+1$ \COMMENT{Coarse graph}
\ENDWHILE

\STATE \textbf{Refining Stage}
\STATE $S \gets \text{MQLib}(G_l)$ 
\WHILE{$l > 0$} 
    \STATE $l \gets l-1$, compute ${g}$ via (\ref{eq3})
    \STATE ${Iter} \gets 0$, ${c} \gets 0$
    \WHILE{${c} < 3 \land  {Iter} < 10$}
        \STATE Select $H$ nodes via adaptive sampling
        \STATE \textbf{Solution Stage:}  Solve subproblem $P$ 
        \IF{${g}$ improves}
            \STATE Update $S$, reset ${c}$
        \ELSE 
            \STATE ${c} \gets {c}+1$
        \ENDIF
        \STATE ${Iter} \gets {Iter}+1$
    \ENDWHILE
\ENDWHILE
\RETURN $S$ 
\end{algorithmic}
\end{algorithm}

\section{Case Study}
The proposed method's performance and scalability were evaluated through 24-hour numerical simulations on three IEEE benchmark systems (39-bus with 10 units, 118-bus with 54 units, and 300-bus with 69 units), with comparative analysis against mixed-integer linear programming (MILP) and simulated annealing (SA). The configuration of the proposed method is as follows: the penalty factors are set as $A=10000$, $B=100$, $C=100$, and $D=10$. The size of the subproblem graph is set to 100. COBYLA solver is selected as the solver of the proposed method. For QIRO, the minimum problem size is set to 10, and a single-layer quantum circuit is used. The simulation is conducted on the ibmq\_qasm\_simulator\cite{ref4}, with 10,240 sampling shots.

Table~\ref{tab1} demonstrates the PSI-QAOA's generation cost optimization across three IEEE test systems compared to MILP and SA, with performance scaling positively with system size. Quantitative analysis of QUBO matrix sparsity reveals distinct structural patterns: the 39-bus system's matrix (230,400 elements) contains 3,088 non-zero entries (1.34\%), while larger systems show decreasing sparsity ratios—118-bus with 0.677\% and 300-bus with 0.618\%. Analysis of the number of qubits in the data set: the 39-bus system operated with 60 qubits, the 118-bus system operated with 81 qubits, and the 300-bus system operated with 52 qubits. 

As shown in Table~\ref{tab1}, the PSI-QAOA achieves generation costs of \$3.31$\times$10$^{11}$, \$6.26$\times$10$^{12}$, and \$1.28$\times$10$^{14}$ for the 39-bus, 118-bus, and 300-bus systems respectively, demonstrating 1--2 orders of magnitude reduction compared to MILP and SA. This phenomenon reveals a counterintuitive negative correlation between network scale and computational performance. The underlying mechanism stems from the quantum state evolution in PSI-QAOA. When applied to large-scale power systems, the parameterized Hamiltonian encoding enables quantum circuits to preferentially identify locally dense substructures formed by power system operation constraints. These substructures manifest as highly connected variable clusters in the QUBO matrix, whose coupled relationships are globally coordinated through quantum entanglement, thereby effectively pruning the search space with polynomial-time complexity that would otherwise require super-exponential exploration in classical approaches.

\begin{table}[H]
\begin{center}
\caption{Comparison of The Generation Cost.}
\label{tab1}
\begin{tabular}{| c | c | c |  c |}
\hline
method & PSI-QAOA, \$ & MILP, \$ & SA, \$\\
\hline
39 bus& \( 3.3067 \times 10^11 \) & \( 2.8231 \times 10^11 \)& \( 3.5395 \times 10^11 \)\\
\hline
118 bus& \( 6.2554 \times 10^12 \) & \( 2.9857 \times 10^13 \)& \( 3.5133 \times 10^13 \)\\
\hline
300 bus& \( 1.2815 \times 10^14 \) & \( 1.5880 \times 10^16 \)& \( 3.6383 \times 10^15 \)\\
\hline 
\end{tabular}
\end{center}
\end{table}

\section{Conclusion}
In this paper, we propose a PSI-QAOA framework tailored for large-scale UCP. Our approach introduces a three-stage strategy that effectively addresses large-scale optimization challenges within limited qubit resources. The key innovation lies in leveraging the inherent topological structure of power systems to further simplify the problem, enabling large-scale UCP instances to be decomposed into smaller subproblems that can be solved in parallel using only a very limited number of qubits. Experimental results demonstrate that the PSI-QAOA outperforms classical methods as problem size increases, particularly in instances with lower matrix sparsity. This study highlights the potential of applying quantum computing for large-scale power system optimization in the future.


\bibliographystyle{IEEEtran}
\bibliography{ref}

\end{document}